\documentstyle[12pt,epsfig]{article}

\begin{document}

\title{Thermodynamic chaos and infinitely many critical exponents in the Baxter-Wu
model}
\author{S.K. Dallakyan$^{1,2}$\thanks{
e-mail: sargis@cerfacs.fr}, N.S. Ananikian$^{1}$\thanks{
e-mail: ananik@jerewan1.yerphi.am} and R.G. Ghulghazaryan$^{1}$\thanks{
e-mail: ghulr@moon.yerphi.am} \\
$^{1}${\normalsize Department of Theoretical Physics, Yerevan Physics
Institute,}\\
{\normalsize Alikhanian Br.2, 375036 Yerevan, Armenia}\\
$^{2}${\normalsize Parallel Algorithms Project, } \\
{\normalsize European Center for Research and } \\
{\normalsize Advanced Training in Scientific Computation, } \\
{\normalsize 42 Avenue Gaspard Coriolis, 31057 Toulouse Cedex, France}}
\maketitle

\begin{abstract}
The mechanisms leading to thermodynamic chaos in the Baxter-Wu model is
considered. We compare the Baxter-Wu model with triangular antiferromagnets
and discuss the difficulties related to the modeling of thermodynamic chaos
by disordered models. We also discuss how to overcome the problem of
infinitely many order parameters. Then we consider the Baxter-Wu model in a
complex magnetic field and show the existence of infinitely many critical
exponents in this model.

\begin{enumerate}
\item[PACS numbers]  05.50.+q; 05.90.+m; 64.60.Cn; 64.60.Fr
\end{enumerate}
\end{abstract}

\section{Introduction}

In contrast to microscopic$/$classical chaos the physics behind
thermodynamic chaos is far from being well understood. Defined to be the
macroscopic state of a system with chaotically broken translational
symmetry, this phenomena is one of the main characteristics of the glassy systems%
\cite{1}. Traditionally these systems investigated by introducing randomness
in the microscopic parameters. Although in many cases this can be physically
justified, it leaves the origin of randomness questionable, especially when
microscopic parameters in turn are measured from macroscopic response of the
system. Recent progress in low temperature physics shows that frustration
and competing interactions are more important for thermodynamic chaos than
the randomness itself\cite{2}. Thus the interest to the deterministic models
of Statistical Mechanics (SM) where frustration and competing interactions
are not the result of randomness.

The antiferromagnetic Ising model on triangular lattices is a classical
example of the model incorporating both frustration and competing
interactions\cite{A1,A2}. In 2d, exact solution in the absence of magnetic
field reveals that this model is not capable to exhibit thermodynamic chaos%
\cite{Wan}. Recently, we investigated the antiferromagnetic Ising model in a
magnetic field on Husimi tree, which being approximation to regular
lattices, allows to preserve frustrative nature of antiferromagnetic
interaction\cite{S!}. Although this model has an interesting phase
structure, no thermodynamic chaos has been found. The other approximations
preserving frustration also show no sign of thermodynamic chaos in
triangular antiferromagnets\cite{S111}. The reason is the following:
frustration on a triangle results macroscopic degeneracy of the ground
state, including chaotic configurations, but due to symmetry of the
Hamiltonian these configurations are not observable at the macroscopic
level. On the other hand, the magnetic field, which competes with
antiferromagnetic interaction completely confines frustrations. Thus a
special symmetry breaking field, other than uniform, is necessary for
observing thermodynamic chaos in tr iangular antiferromagnets.

The Baxter-Wu model in a magnetic field is one of the models where
frustration and competing interactions present simultaneously\cite{Bax}. The
thermodynamic chaos found in this model by Monroe\cite{A3} has been
investigated in some details by the present authors\cite{A4,A5}. Universal
transition to chaos is one of the interesting phenomena which has been
found. The main difficulties which one faces is that it is necessary to
introduce infinitely many order parameters to describe the system in a
chaotic phase. It is interesting to note that similar problem exist also in
disordered models of spin glasses, namely one needs to introduce Parisi's
order parameter functional or local Edwards-Anderson order parameters in
order to describe the system in a glassy phase\cite{Glass}.

Fortunately, for the deterministic models one can use the thermodynamic
formalism of dynamical systems to infer the universal characteristics in
chaotic phases. In particular, distribution of the local Lyapunov exponents
shows universal scaling behavior, similar to one found for a logistic map%
\cite{A4,A5}. Although similar universalities have been also found in the
distribution of the local Edwards-Anderson order parameters of disordered
models\cite{Br}, as we shall see, in many cases disordered models are not
capable to model thermodynamic chaos.

In this paper we investigate the Baxter-Wu model in a complex magnetic
field. At this point, it is interesting to draw some parallels between
Quantum Mechanics$/$Quantum Field Theory (QFT) and SM. Investigation of the
singularities of the scattering matrix in a complex energy$/$momentum plane
reveals an information about bounded states and resonants, which otherwise
is difficult to obtain\cite{Qu}. But in contrast to scattering matrix,
singularities of thermodynamic quantities are not necessarily isolated. In
fact, as it has been proven by Lee and Yang in 1952\cite{Lee}, zeros of the
partition function of the 2d Ising model in a complex magnetic field lie on
a curve, which tend to the real axis at the phase transition point and
become dense in the thermodynamic limit. Soon after, Huang in his seminal
textbook of SM\cite{Huang} rises a question about possibility of having SM
model, zeros of the partition function of which pinch the real axis at some
range instead of a single point.

Recently, we investigated the Baxter-Wu model in a complex temperature plane%
\cite{S2}. It has been shown that the Fisher zeros of this model densely
fills a domain near the real axis. As we shell see, the Lee-Yang zeros of
the Baxter-Wu model satisfy the criterion mentioned by Huang. This allows us
to prove the existence of infinitely many critical exponents in this model.

The paper is organized as follows. In Section 2 after introducing the
Baxter-Wu model we discuss principal differences in modeling thermodynamic
chaos by deterministic and disordered models. In Section 3 we present our
numerical results for the phase structure of the Baxter-Wu model in a
complex magnetic field and discuss the role of thermodynamic chaos on it. In
Section 4 we present our conclusions.

\section{The Baxter-Wu model}

The Hamiltonian of the Baxter-Wu model in a magnetic field has the following
form:

\begin{equation}
{\em H}=-J_{3}\sum_{\triangle }\sigma _{i}\sigma _{j}\sigma _{k}-h\sum
\sigma _{i}  \label{0}
\end{equation}
where $\sigma _{i}\in \{-1;+1\}$ are Ising variables, $J_{3}$ and $h$ are
three-site interaction strength and magnetic field respectively. The first
sum goes over all triangles and the second one over all sites. Like two-site
interacting Ising model this model has multiple applications and is one of
the few models of SM exactly solvable 2d. For the first time the solution
has been found by Baxter and Wu in 1973 by using Bethe Ansatz method\cite
{Bax}. Later, it has been shown that this solution is a particular case of
more general solution of the eight-vertex model\cite{Bax1,Bax2}. Other
interesting relations between Baxter-Wu model and other exactly solvable SM
models can be obtained using generalized star-triangle relations\cite
{Bax2,Bax3}. Recently, the solution by the Bethe Ansatz method has been
generalized to include more general boundary conditions and an interesting
conjecture in the framework of Conformal Field Theory has been proposed,
viz. that the Baxter-Wu model and 4-state Potts model share the same
operator contents\cite{wu}.

On a single triangle the ground state of the three-site interaction consist
of configuration where all spins aligned at the same direction (up or down
depending on the sigh of $J_{3}$) and of configurations obtained from this
one by reserving the spins at arbitrary two sites of the triangle. This
makes the ground state of the Baxter-Wu model highly degenerate. In
particular, an arbitrary alignment of the spins along arbitrary direction
can be achieved starting from the uniform configuration without altering the
total energy of the system. By encoding these configurations in binary
sequences it becomes clear that the ground state of the Baxter-Wu model
involve uniform, modulated, as well as chaotic configurations. Note that
third spin in the Hamiltonian of Baxter-Wu model acts like a (pseudo) random
two-site interaction strength.

As we mentioned in the introduction, the macroscopic degeneracy itself is
not sufficient for thermodynamic chaos. In addition to macroscopic
degeneracy a symmetry breaking field is necessary, which will pick up a
particular chaotic configuration of the ground state when averaging over all
configurations. As we shell see, in the Baxter-Wu model the magnetic filed
with opposite sign to $J_{3}$ satisfies this criteria.

To that end, let us consider the Baxter-Wu model on the Husimi tree\cite{A3}%
, so that frustrative nature of the three-site interaction can be preserved
and at the same time analytical expression for thermodynamic quantities can
be obtained, even in the presence of magnetic field.

The magnetization at the central site of the Husimi tree as a function of
temperature $T,$ magnetic field $h$ and three-site coupling $J_{3}$ is given
by\cite{A3}

\begin{equation}
m_{n}(T,h,J_{3})=\frac{\mu x_{n}^{\gamma }-1}{\mu x_{n}^{\gamma }+1}
\label{1}
\end{equation}
where $n$ numbers the generation in the hierachies of Husimi trees, $\gamma $
is equal to the twice the coordination number and $x_{n}$ is given by the
following recurrent relation

\begin{equation}
x_{n}=f(x_{n-1}),\qquad f(x)=\frac{z{\mu }^{2}x^{2(\gamma -1)}+2\mu
x^{\gamma -1}+z}{{\mu }^{2}x^{2(\gamma -1)}+2z\mu x^{\gamma -1}+1}  \label{2}
\end{equation}
where $z=e^{2J_{3}/kT}$ and $\mu =e^{2h/kT}$. Initial condition for the
recurrent relation (\ref{2}) depends on the boundary conditions (e.g. $%
x_{0}=1$ corresponds to the free boundary condition).

It is interesting to point out that the recurrent relation (\ref{2}) enters
also in the formulas for the expectation values of quantum mechanical
operators in some field theoretical models\cite{S3,jo}.

Depending on $T,h$ and $J_{3}$ the attractor of the map consist of a stable
point, periodic cycles or strange attractors, so that our system is in
uniform (paramagnetic of ferromagnetic), modulated or glassy phases
respectively (Figure 1).

Note that as far as the dynamic of the map (\ref{2}) is concerned there is
no difference whether it describes the dynamic of microscopic quantities or
the distribution of macroscopic one. The only difference is in the
interpretation of the results, so that in the former case it describes time
evolution of microscopic quantities, whereas in the later case site to site
variation of macroscopic quantities. For instance, the map which is involved
in the formulas for magnetization of the ANNNI model on the Bethe lattice
natural arises in the microscopic dynamic of a neuron with non-monotonic
transfer function\cite{net}.

Taking into account the discussion in the introduction, one can see that the
large variety of phases, which are typical in low temperature physics and
biophysics are the result of collective effects of frustrations {\em and}
competing interactions.

The main difficulties lie in the parameter space where the system exhibits
thermodynamic chaos. In the uniform or modulated phases one can use
conventional order parameters, such as magnetization, but at every
bifurcation point the system undergoes a continuous phase transition and
they number doubles. Thus in a chaotic phase we end up with infinitely many
order parameters.

To find a possible solution to this problem one can use an invariant measure 
$P(m)$ of the map (\ref{2}) to define e.g. $q=\int mP(m)$ as an order
parameter. But, as it is well known, like in axiomatic QFT or in stochastic
dynamical systems, in general, even the first few momentums are not
sufficient for describing the system. Thus we have to use the thermodynamic
formalism of dynamical systems for the complete description of the system%
\cite{R12,RR12}.

Let us recall that using SM\ one anticipates to get rid of the large amount
microscopic degrees of freedom and to avoid the solution of the complicated
microscopic dynamics involving thermostat. The thermodynamic formalism of
dynamical systems, on the other hand, allows one to obtain the quantities
which describes the dynamical systems, e.g. the spectrum of Lyapunov
exponents or generalized dimensions\cite{Mul}. By applying thermodynamic
formalism of dynamical systems we do not anticipate to get rid of infinitely
many order parameters. Instead, we obtain an information about our system in
terms of Lyapunov exponents or generalized dimensions. That is the
universalities in the distribution of these quantities which allows to avoid
the measurement of infinitely many order parameters\cite{A5,AA5}.

To compare deterministic and disordered models of thermodynamic chaos let us
consider the problem of modeling thermodynamic chaos in a computer. In order
to distinguish the modulated phase with large period from a chaotic one we
have to simulate the deterministic model on a very large lattice. On the
other hand, we can simulate this system on a small lattice by a disordered
model with appropriately chosen measure for random parameters. But since
strange attractors of dynamical systems with a few exceptions (e.g.
hyperbolic dynamical systems\cite{RR12}) have infinitely many invariant
measures\cite{Rue}, disordered models are not capable to model thermodynamic
chaos at all temperatures and boundary conditions.

It is interesting to note that thermodynamic or special chaos can be
observed in deterministic systems by changing boundary conditions only\cite
{Cha1,Ku}.

\section{Lee-Yang singularities}

An attractive feature of disordered models is that it is believed that only
a few set of critical exponents are needed to describe different phase
transitions taking place in spin glasses\cite{Glass}. Complex temperature$/$%
field analysis is one of the tools used among many others to extract the
critical exponents\cite{Co1,Co2}.

Zeros of the partition function in the complex temperature$/$field plane
coincide with the Julia set of renormalization group map and provide
information about phase transition points and critical exponents\cite{R9,R2}%
. In particular, the density on the curve on which partition function zeros
lie and the angle which this curve make with the real axis are directly
related to the critical exponents.

The fact that the system is in complex temperature and$/$or magnetic field
does not mean that this system is unphysical or nonunitary. In fact, there
are well know examples in the literature where duality or star-triangle
relation map the system with real temperature and magnetic field into a
system with complex temperature and$/$or magnetic field.

Zeros of the partition function of the Baxter-Wu model on the Husimi tree
satisfy the following relation\cite{S2}

\begin{equation}
\mu x^{\gamma }+1=0  \label{3}
\end{equation}

on the attractor of the map (\ref{2}).

In Figure 2 we plot the zeros of the partition function in a complex field
for different temperatures. One can see at Figure 2a that zeros of the
partition function approach to only a few set of critical points located on
the real axis, whereas at low temperatures (Figures 2b) the Lee-Yang
singularities densely fill domains near real axis which include patches of
the real axis. This indicates a condensation of the phase transition points.
In these domains there are infinitely many different ways leading to a given
point in the real axis, which shows the existence of infinitely many
critical exponents in the Baxter-Wu model. The existence of infinitely many
critical exponents can be seen also from Figure 1. Since at every
bifurcation point one can define critical exponents in terms of staggered
magnetizations, in a chaotic phase there are infinitely many critical
exponents.

We reiterate that in a real experiment there is no need to measure neither
infinitely many order parameters nor infinitely many critical exponents.

\section{Conclusion}

In this work we studied thermodynamic chaos in the Baxter-Wu model. Although
Baxter-Wu model shares many features of triangular antiferromagnets, the
external magnetic field competes with the three-site interaction leaving the
ground state highly degenerate, a phenomena which makes chaos observable at
the macroscopic level.

This phenomena is well known in disordered models of spin glasses, where
similar result can be obtained by introducing randomness in the microscopic
parameters (e.g. magnetic field or interaction strength). Nevertheless,
disordered models, in general, fail to model thermodynamic chaos, since
probabilistic measure (which in disordered models we have to chose
phenomenologically) is not unique and varies depending on temperature and
boundary conditions.

Investigation of the Lee-Yang singularities revealed the existence of
infinitely many critical exponents in the Baxter-Wu model. The problem with
infinitely many order parameters and critical exponents is that in a real
experiment one have to perform infinitely many measurements. Notice that a
similar problem exist also in nonrenormalizable QFT, where nonrenormalizable
interaction leads to infinitely many counter terms and corresponding
coupling constants\cite{Cheng}. As we mentioned, for the Baxter-Wu model we
can use the thermodynamic formalism of dynamical systems, which allows us to
completely describe the systems in chaotic phases by a few set of universal
quantities (e.g. the slope in the distribution of the Lyapunov exponents\cite
{A5,AA5}). We hope that, duality relation between Baxter-Wu model and $Z(2)$
gauge symmetric model involving nonrenormalizable three-plaquette
interaction, which we recently found\cite{S3}, will help us to understand
the connection between nonrenormalizable QFT and thermodynamic chaos.

\section{Acknowledgments}

We thank B. Hu for the hospitality at the Centre for Nonlinear Studies at
Hong Kong Baptist University. SKD thanks D. Shepelyansky for helpful
discussion.This work is supported in part by grants INTAS-97-347 and A-102
ISTC project.

\begin{figure}[b]
\epsfysize=13cm\centerline{\epsfbox{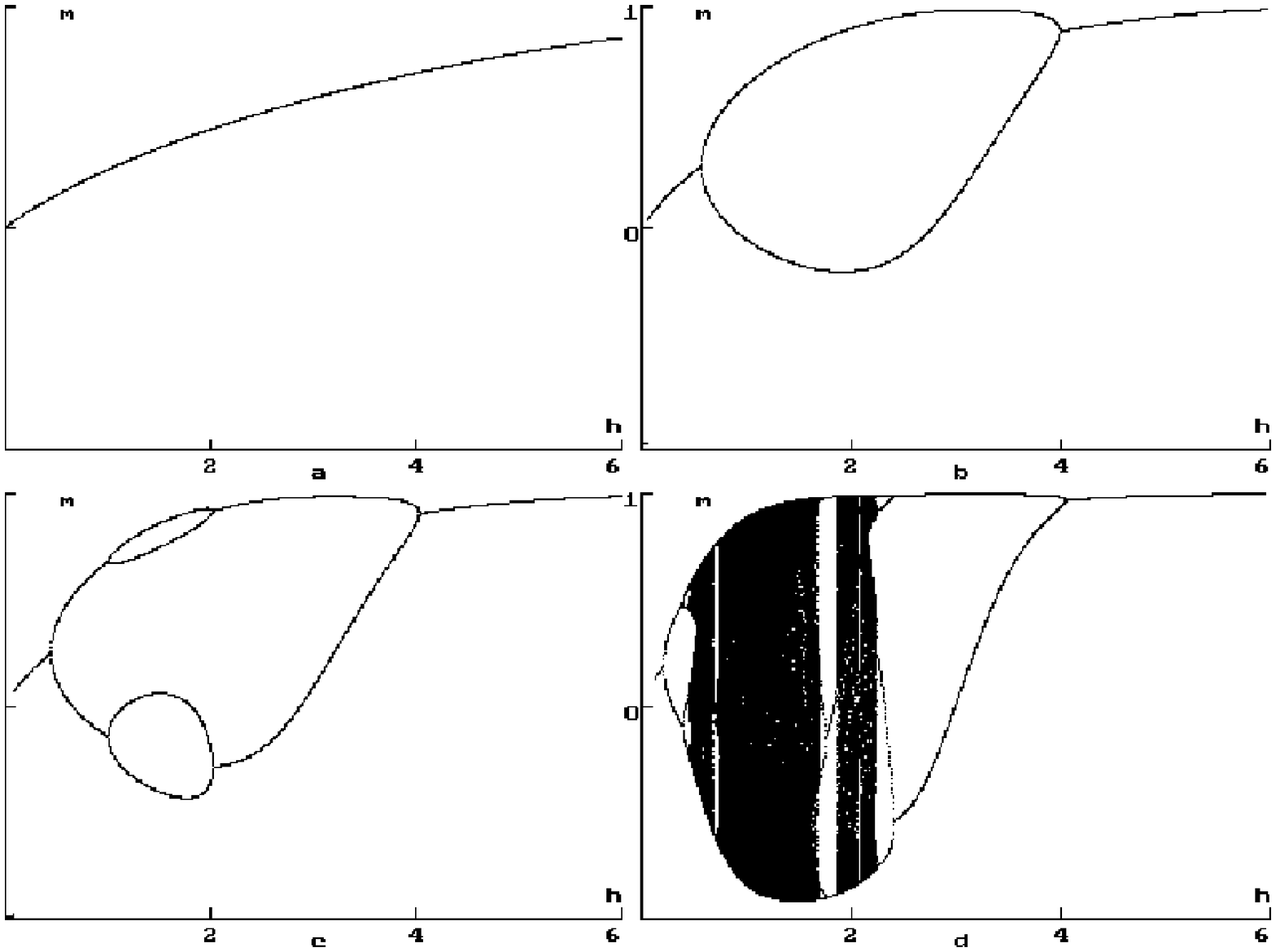}}
\caption{Plots of $m$ versus $h$ for different temperatures $T$ ($\gamma =4,
J_{3}=-1$). a - $T=3$, b - $T=1.3$, c - $T=1.15$, d - $T=0.7$. }
\label{figtwo}
\end{figure}
\newpage

\begin{figure}[b]
\vspace{-3cm} 
\begin{tabular}{c}
\epsfig{file=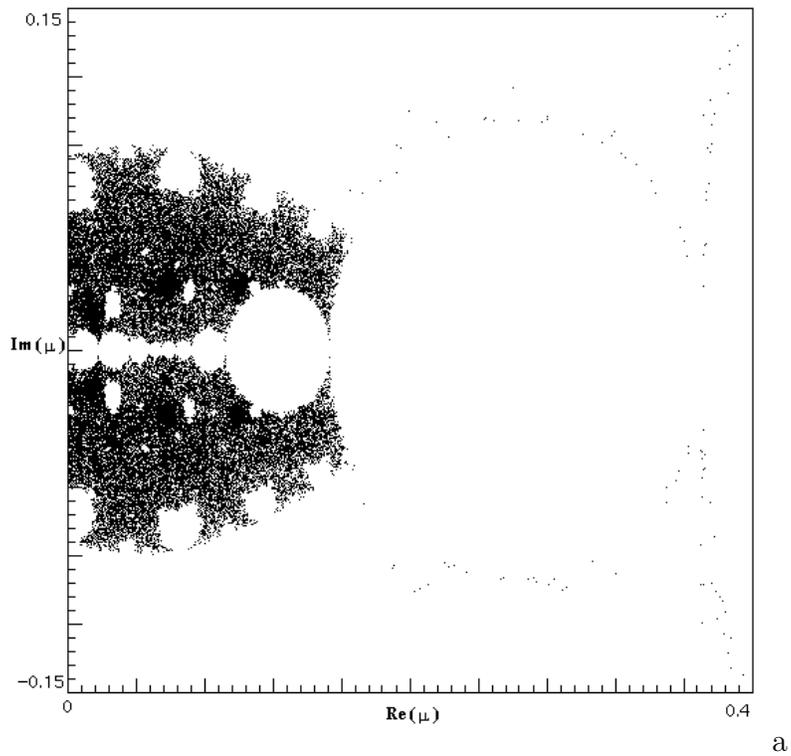,width=10cm,height=10cm} a \\ 
\epsfig{file=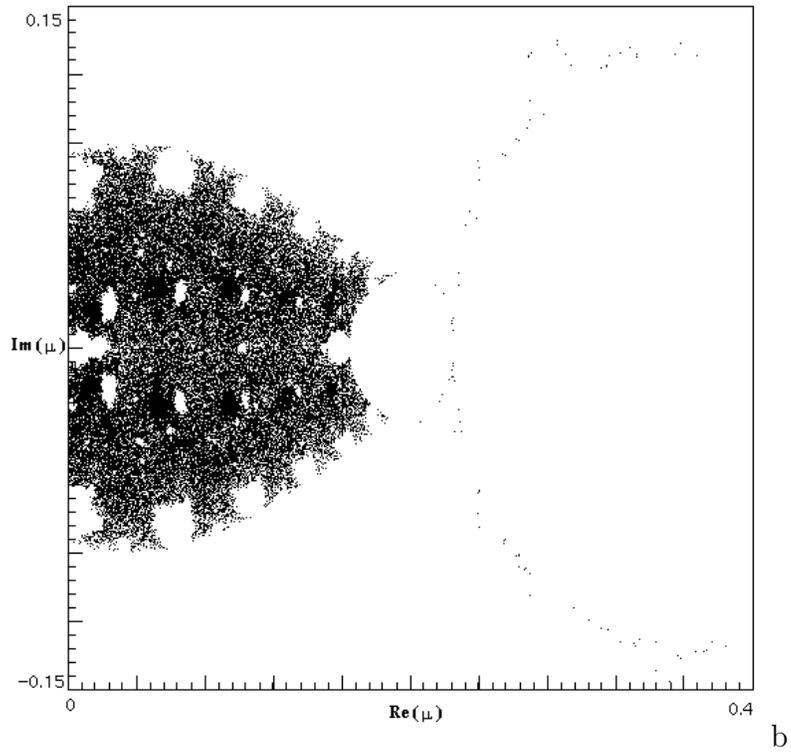,width=10cm,height=10cm} b
\end{tabular}
\caption{Lee-Yang singularities of the Baxter-Wu model ($\gamma $ = 4). a - $%
z=6.05$, b - $z=7.0$.}
\label{figthree}
\end{figure}

\end{document}